# 4

# Biomechanical modeling of the human tongue

Maxime C<small>ALKA</small>[1,3], Pascal P<small>ERRIER</small>[2],
Michel R<small>OCHETTE</small>[3] and Yohan P<small>AYAN</small>[1]

[1] *TIMC, Université Grenoble Alpes, Grenoble, France*
[2] *GIPSA-lab, Université Grenoble Alpes, Grenoble, France*
[3] *Ansys, Villeurbanne, France*

## 4.1. Introduction

The tongue is a crucial organ for performing basic biological functions, such as chewing, swallowing and phonation. Understanding how it behaves, its motor control and involvement in the execution of these different tasks is therefore an important issue for the management and therapeutic treatment of pathologies relating to these essential functions so that quality of life can be preserved. Biomechanical modeling of this organ is one of the key steps towards this understanding, and it will be an important tool to predict and control the functional impact of lingual surgery in the field of computational assistance to medical-surgical methods.

One of the specificities and difficulties of the biomechanical modeling of this lingual articulator is related to the fact that it is a muscular hydrostat (eight even muscles and one odd muscle) that deforms very significantly (Napadow *et al.* 1999) through the action of muscles that form either entirely or very largely an integral part of the tongue, which makes it possible to obtain complex and varied shapes. Added to this is the fact that when it comes to both speech production and swallowing, the dynamics of tongue movements are crucial elements for the motor tasks' success (effective communication for speech, transition from the food bolus to the esophagus when swallowing). This implies that tongue modeling should account for the constraints that are induced by the elastic properties of tissues on the temporality of lingual gestures. These observations led us to propose nonlinear elastic modeling of tissues (hyperelasticity) and to study the dynamic behavior of the finite element model through a transient analysis with damping. This type of analysis, relatively rarely used in the field of tissue biomechanics, is important for the lingual organ, which has the particularity of moving and deforming at very significant speeds for dynamics equations (10 to 20cm/s). A research perspective not yet explored is to implement a visco-hyperelastic law which results from experimental tests on lingual tissues (Karkhaneh *et al.* 2018).

In the rest of this chapter, we will first present some notions on the anatomy of the tongue and the vocal tract, which are necessary to understand the model structure. This will be followed by a rapid look at the different models of the tongue which allow us to understand the framework of our work and the hypotheses that underlie our modeling approach. The models' implementation will then be described in three phases:

1) geometry and finite element mesh ;

2) the behavioral laws of passive soft tissues and lingual muscle structures ;

3) the mechanical and temporal parameterization of our numerical simulations.

The results from the simulation of the impact of different muscles on the shape of the tongue will then be presented and analyzed, particularly in relation to the results published in literature. To conclude this chapter, we will present the prospects that this model offers for future work.

## 4.2. Anatomy of the tongue: environment, topology and partitioning

The tongue is rooted in the pharyngeal cavity on a small movable bone located just above the larynx, the hyoid bone, and it grows forward to the teeth (Figure 4.1). Occupying most of the oral cavity, it is located above the floor of the mouth and is directly attached to the mandible in its anterior part, at the level of the apophyses geni at the height of the chin, and indirectly, *via* the muscles of the base of the mouth, to the body of the mandible, in its anterior and lateral part. It is also connected by extrinsic muscles to the styloid process and soft palate, as well as to the pharynx *via* the upper and middle pharyngeal constrictor muscles, and by the mucosa.

Ovoid in shape, when it is inert, with a length of about 10 cm, a maximum width of 5 cm and a mass varying from 60 to 80 g, the tongue is a musculomanous organ covered with mucous membranes, where the taste buds are located. It is separated vertically by a fibrous lamina, the lingual septum, into two symmetrical halves with respect to the median sagittal plane of the head.

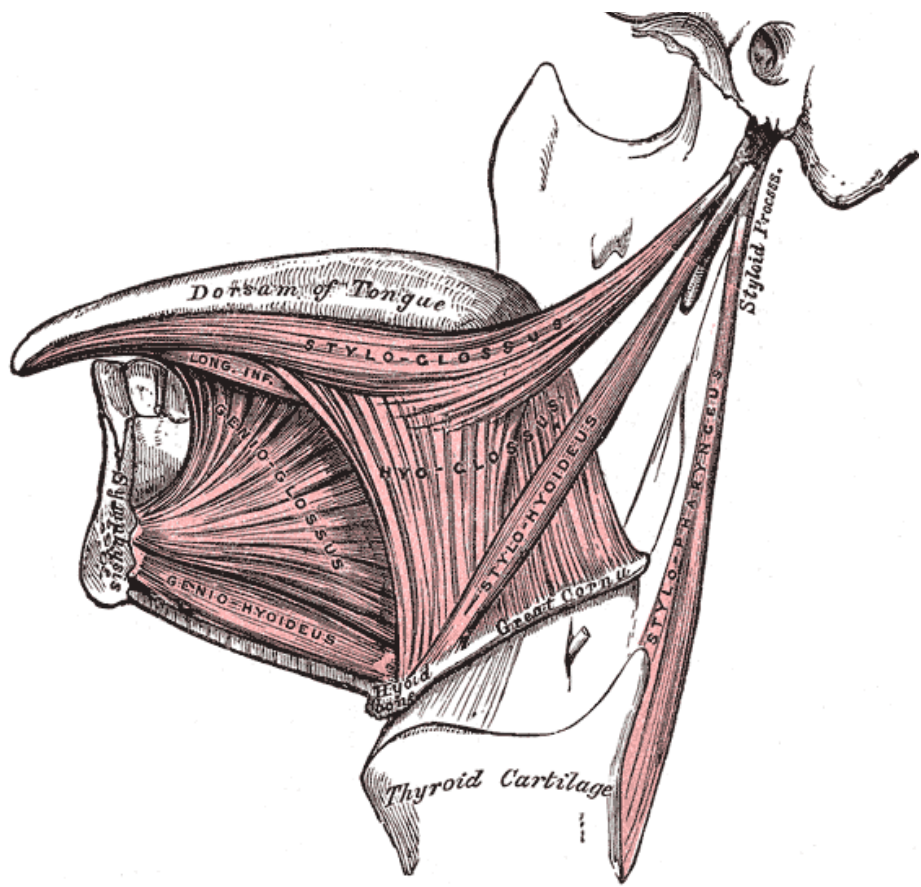

**Figure 4.1.** *Main muscles of the tongue. (Gray 1918)*

From a muscular point of view, the tongue is an extremely complex organ given its large number of muscles, how they interlace and their geometry (Figure 4.1). It is subject to the direct action of eight pairs of muscles: the genioglossus divided into three parts – anterior, medium, posterior –, the styloglossus, the hyoglossus, the inferior and superior longitudinalis and the verticalis, as well as a muscle called the transversalis muscle. They are traditionally divided into two categories: extrinsic muscles, originating outside the tongue and allowing control of the overall movements of the tongue (protrusion, retraction and depression), and intrinsic muscles, internal to the tongue, which do not have insertion on bone structures and which control changes in finer shape.

## 4.3. State of the art on the biomechanical modeling of the human tongue

As mentioned in the introduction, the biomechanical modeling of the human tongue is a complex but important subject to help understand the mechanical behavior and functions of this organ. Even if they are not numerous, different numerical models have been developed for the study of speech production, or, in a clinical context, to better understand swallowing or sleep apnea phenomena.

The first three-dimensional model of the tongue adopting the finite element method (FE) was developed by Kakita and Fujimura (1977) and used to quantify, for the first time, the role of the intrinsic and extrinsic muscles of the tongue in speech production. The tissues were then modeled by a linear constitutive law.

Wilhelms-Tricarico (1995) developed a more sophisticated 3D FE model, with a constitutive law of hyperelastic soft tissue and linear viscosity. In this model, the stress induced by muscle contraction is calculated as the sum of the active and passive stresses of the muscle. The active stress is influenced by the direction of the fibers and by a time-dependent parameter representing the level of muscle activation.

In parallel, Payan and Perrier (1997) worked on a two-dimensional FE model, choosing a compromise between realism and computational time to study the control of speech production. This model has made it possible to simulate a large number of articulatory movements, which have been compared with the movements of the tongue recorded by electromagnetometry on human speakers, and which have enabled the development of a model for controlling speech production (Patri *et al.* 2015).

In 2004, Dang and Honda (2004) offered a 2.5D FE model. The specificity of the mesh that is proposed in this model is to only represent a reduced part of the tongue around the median sagittal plane, inserted into the oral cavity (mandible, larynx, lips, hard palate, velum). Muscles are represented by active stresses that follow the principles of Hill's muscle model (Winters 1990), and are superimposed on the passive stresses of the finite element structure. These active stresses are applied in the directions defined by macrofibers that represent the main directions of muscle fibers. The rest of the soft tissues are modeled by an isotropic Hooke's law. This model is not sufficient to reproduce the forms of the tongue that are observed in pathologies. Fujita *et al.* (2007) have improved it to make it usable in a clinical context with a tongue represented in 3D in its entirety. This model retains the mechanical, passive and active properties of the 2004 model, but has yielded encouraging results regarding the possibility of using it for surgical planning assistance. Finally, more recent work carried out in 2009 by the same team (Fang *et al.* 2009) refined the model and its control to study the muscular recruitment required to produce Japanese vowels.

After Wilhelms-Tricarico (1995), several three-dimensional models will use hyperelastic laws of behavior (Rodrigues *et al.* 2001; Gerard *et al.* 2006; Buchaillard *et al.* 2007, 2009; Stavness *et al.* 2012; Wang *et al.* 2013).

In 2006, Gerard *et al.* (2006) proposed a 3D FE model of the tongue including a Yeoh-like hyperelastic distribution, with a tissue incompressibility hypothesis. The parameters of this law were obtained by inverse methods after experimental tests carried out by indentation on a fresh cadaver tongue (Gerard *et al.* 2005). Since then, in *vivo* aspirations techniques have also been used to estimate lingual elasticity in small deformations (Schiavone *et al.* 2008; Kappert *et al.* 2021). Buchaillard *et al.* (2007, 2009) subsequently improved this FE model, in particular by modifying the coefficients of Yeoh's law to stiffen lingual tissues, in order to account for the intrinsic nature of living tissues compared to those of a cadaver. Muscle actions were represented by macrofibers exerting forces on the passive matrix, and these forces were generated according to the principles of the motor control model from Feldman (1986). The biomechanical model of Buchaillard *et al.* (2007, 2009) made it possible to refine the characterization of the action of lingual muscles, to study the impact of gravity, and to propose the first simulations of the impact of cancer surgery on lingual mobility.

The finite element mesh and geometric modeling of muscle structures proposed by Buchaillard and colleagues formed the basis of the FE model of tongues which was developed by Sydney Fels' team at UBC Vancouver, in the context of the *open source platform* ArtiSynth®, which was at the heart of the work of Stavness *et al.* (2011) and many others thereafter (Stavness *et al.* 2012; Gick *et al.* 2017; Anderson *et al.* 2019). The tongue is coupled with the palate and mandible. The muscles are represented by active elements implementing Hill's model (Winters 1990). More recently, Kappert *et al.* (2019) used this FE tongue model to create a surgical simulation tool that includes postoperative suture modeling, with the aim of predicting the functional consequences of partial glossectomies.

Pelteret and Reddy (2012) performed a 3D FE tongue model, coupled with a neural model of excitation, which they used to assess the ability of the tongue to maintain its posture in the presence of aerodynamic pressure generated by airflow in the context of sleep apnea. In this FE model, a simple constitutive law of the Neo-Hooke type characterizes soft tissues. The complexity of this model lies in the muscle contraction represented by a Hill model controlled by a neural system, which replaces the ad hoc scalar constant controlling with the force amplitude found in the other models. The neuronal activations of the control system are determined for each muscle group using a genetic algorithm.

Wang *et al.* (2013) developed a 3D model of the tongue, with the aim of determining the best placement of sensors used in electromagnetic articulography (EMA) on the surface of the tongue to record lingual displacements. It is a rather dense FE model, consisting of 120 nodes and 64 elements. A two-parameter Mooney-Rivlin constitutive law is used to represent the tissues of the tongue. The muscle model is based on that of Blemker *et al.* (2005).

Recently, the model of Gerard and Buchaillard has been improved in the work of Rohan *et al.* (2017) and Hermant *et al.* (2017), which refined mesh topology by segmenting new very high-definition MRI images, increasing the number of model elements and modeling muscles with active elements based on the principles of the Hill model. The model presented in the remainder of this chapter is a continuation of this work.

### 4.4. Our finite element 3D model of the human tongue

#### 4.4.1. *Geometry and mesh*

As said above, the tongue mesh we currently use is derived from that which is successively refined (Gerard *et al.* 2005, 2006; Buchaillard *et al.* 2007, 2009; Hermant *et al.* 2017; Rohan *et al.* 2017). It reproduces the geometry of the tongue extracted by segmenting a high-resolution MRI of a reference subject, for which we have a large amount of kinematic data in speech production and swallowing (Figure 4.2).

This mesh (Figure 4.2), composed of 58,526 nodes and 40,102 tetrahedral and quadratic elements, forms the basis of a finite element (FE) model of the tongue, the characteristics of which will be detailed later in the chapter.

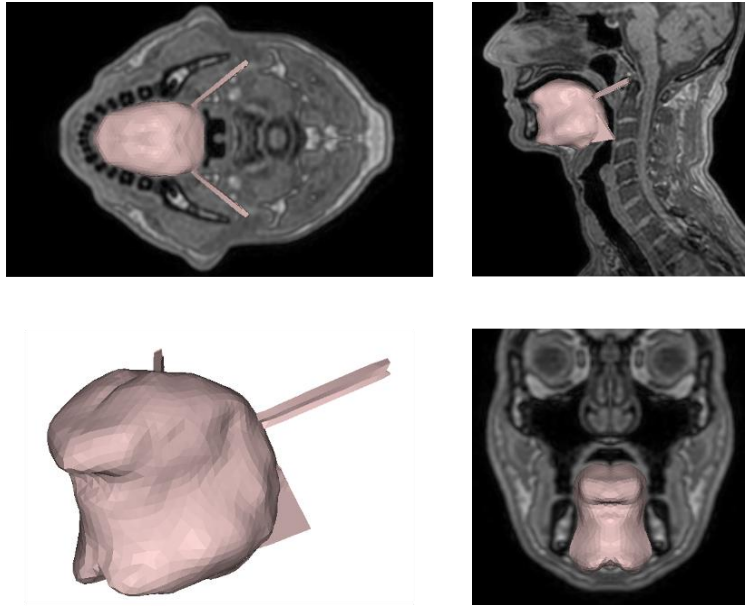

**Figure 4.2.** *3D tongue mesh model inserted in the MRI volume of the reference subject.*

### 4.4.2. Constitutive laws

#### 4.4.2.1. *Passive constitutive laws of lingual soft tissues*

The soft and muscular tissues of the human body are materials with complex mechanical behavior and strong nonlinearities. For the tongue, whose tissues undergo very large deformations during speech production (Napadow *et al.* 1999), Gerard *et al.* (2005) proposed a law of hyperelastic behaviour developed from experimental indentation data collected from a fresh cadaver.

In this model, the behavior of the soft tissues of the tongue is characterized by the following hyperelastic Yeoh distribution:

$$\Psi = C_{10}(I_1 - 3) + C_{20}(I_1 - 3)^2 + \frac{\kappa}{2}(J - 1)^2 \qquad [4.1]$$

with:

$$\kappa = \frac{2C_{10}}{1 - 2\nu}$$

where $\Psi$ is the law of strain energy density, $I_1$ is the first invariant of the expansion tensor (right Cauchy-Green strain tensor), $\nu$ is the Poisson coefficient and $J$ is the Jacobian of the transformation (determinant of the strain gradient tensor).

Through an optimization process using "synthesis analysis" methodology applied to an FE model of a tongue sample similar to that used during indentation, Gerard *et al.* (2005) proposed to retain $C_{10} = 192$ Pa and $C_{20} = 90$ Pa, with a Poisson coefficient $\nu = 0{,}49$ (soft tissues being considered almost incompressible). These values were used in the model simulations presented in Section 4.5. In addition, as the tongue is mainly composed of water, a unit density was chosen for these dynamic simulations.

### 4.4.2.2. *Active Constitutive Law: muscle model*

The anatomical implementation of muscles in the FE mesh of our model consists of defining mesh regions (a set of elements associated with each modeled muscle) in which the constitutive law integrates an active component, in addition to the passive composition of the entire mesh.

The different muscle structures proposed below (Figure 4.3) have been defined on the basis of the most recent anatomical knowledge available in literature (see Acland 2003). The finite elements associated with each muscle are able to deform in response to a stress whose principal direction is specified locally by a vector corresponding to the main direction of contraction of the muscle fibers.

The muscle model implemented was proposed by Nazari *et al.* (2011, 2013). It takes into account the passive elastic properties and active force-generating mechanisms of a muscle, describing an active element as a volume in which muscle fibers are embedded in a matrix of surrounding passive tissues. The main force generation principle is Hill's activation model (Winters 1990) based on Blemker's formulation. *et al.* (2005). This model incorporates a law of variation of active constraint as a function of speed, which functionally accounts for the behaviour of actin-myosin bridges in sarcomeres (Huxley 1957).

A muscle is then modeled as a transversely isotropic material with passive isotropic behavior (section 4.4.2.1), to which an active behavior is superimposed in the direction of the fibers. Thus, the total stress in the direction of the fiber is equal to the sum of the stress generated by passive tissues and the stress due to a contractile element.

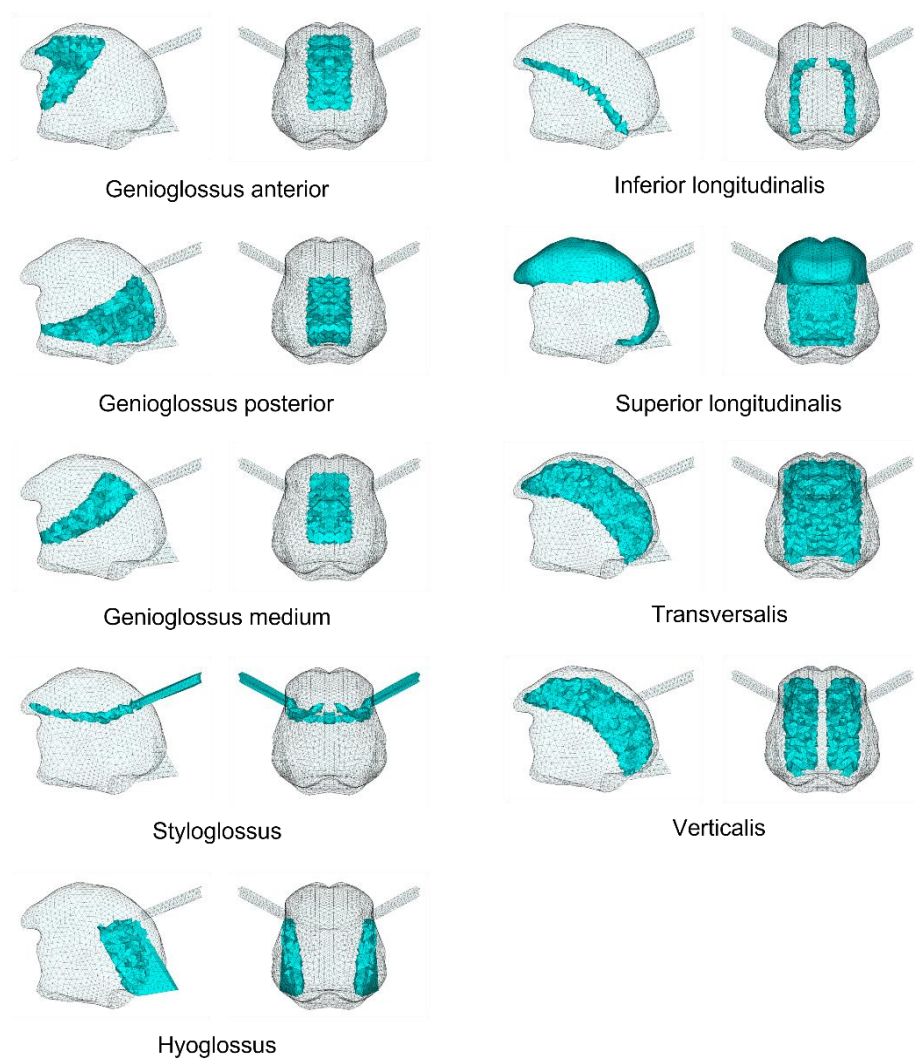

**Figure 4.3.** *Topological description of the different muscles of the tongue. Left: View from the left side. Right: Front view.*

### *4.4.3. Boundary conditions*

#### *4.4.3.1. Tongue/mandible/palate*

As mentioned in the section on the anatomy of the tongue (section 4.2), the tongue is attached to the mandible by extrinsic muscles: the genioglossus, geniohyoid and mylohyoid. A fixed type of contact (no separation or possible slippage between the two meshes), with complete connection between the nodes of the FE mesh of the tongue and those of the mandible, was therefore defined in the regions concerned by these muscles. The tongue can also come into contact with the mandible, palate or teeth in several cases, such as the pronunciation of certain phonemes (for example, the vowel /i/ or the consonant /t/). It was therefore important to model this interaction using sliding contact with separation, which will be considered as ideal in our case.

#### 4.4.3.2. *Hyoid bone/tongue*

The tongue is also attached to the hyoid bone, a small U-shaped bone located behind the tongue, at the bottom of the pharyngeal cavity, above the larynx. This bone has the particularity of not being articulated with any other bone in the human body, which allows it to have great mobility. The tongue is connected to this bone at the posterior base of its root by the hyoglossus muscle and, indirectly, by the geniohyoid and mylohyoid muscles of the floor of the mouth. The contact is modeled by fixed contact.

### 4.5. Numerical simulations

#### 4.5.1. *Transient simulations*

As stated at the beginning of this chapter, the tongue is an organ involved in three tasks, for which not only the final position of the tongue, but also the trajectory it follows to arrive at this position are crucial. In addition, the tongue is an organ that moves very quickly (10 to 20 cm/s), which implies taking into account parameters such as stiffness and damping factors that characterize the dynamics of the system.

For temporal simulations to account for these dynamic properties, transient simulations (simulations resulting from the resolution of Lagrange's equations) must be considered in order to realistically simulate the movements of the tongue. Our model therefore includes a functional consideration of the influence of velocity on tissue deformation *via* Rayleigh damping, whose parameters are $\alpha = 20.0 s^{-1}$ and $\beta = 0.0 s$. The simulations presented below (sections 4.5.3 and 4.5.4) were performed with Ansys Mechanical® software in implicit formulation. In these simulations, we will neglect gravity stresses, as well as residual stresses (Buchaillard *et al.* 2009).

#### 4.5.2. *Temporal activations of tongue muscles*

The temporal course of muscle activation functions during movement is the subject of debate in the field of study of the control of precise motor tasks in humans. The theoretical issues associated with these debates are well illustrated by the confrontation between Gomi and Kawato (1996) and Gribble *et al.* (1997) on the control signals underlying arm movements, and can be summarized by the question of whether the human central nervous system involves a dynamic reversal of the peripheral motor system to control precise movements. Gribble's work *et al.* (1997) showed that precise biomechanical modeling of the peripheral motor system makes it possible to obtain complex trajectories of the motor effectors, without involving dynamic inversion and without involving complex trajectories of control signals. We are part of this perspective, and our work on realistic biomechanical modeling of vocal tract articulators, and the tongue in particular, has validated us in this theoretical choice. We were thus able to show that complex, loop-shaped trajectories observed in speech sequences of the type /igi/, /iga/ or /aga/ could be obtained from variations in constant-speed commands between target commands associated with different phonemes, if they are applied to a realistic biomechanical model of the tongue (Perrier *et al.* 2003). Similarly, complex velocity profiles (temporal variations of tangential velocity) (Payan and Perrier 1997), and the experimentally observed relationship between trajectory curvature and tangential velocity amplitude (Perrier and Fuchs 2008), could be generated using the same modeling approach. All these studies tend to show that the trajectories of speech articulators are strongly influenced by the biomechanical characteristics of the peripheral orofacial system that allows for the production of speech.

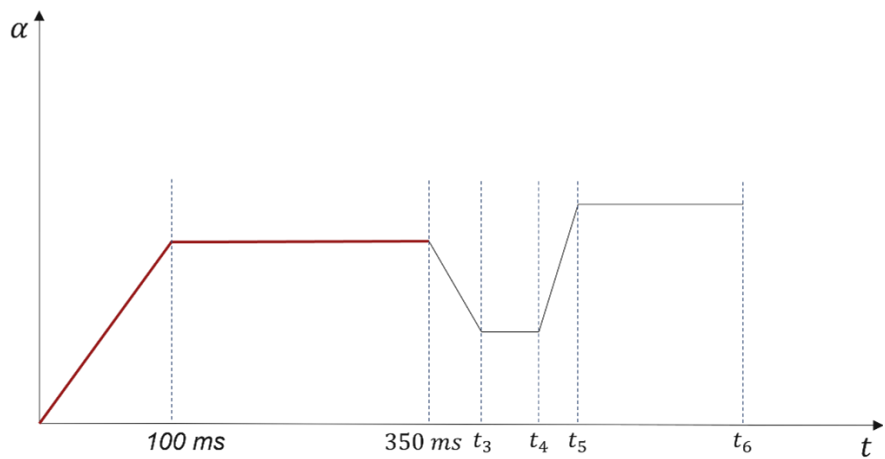

**Figure 4.4.** *Example of a temporal activation sequence. In red, the temporal activity pattern of the muscles shown in Figure 4.3. These diagrams are organized around the notion of transition (increasing or decreasing slope) between target postures (plateaus).*

Therefore, as part of this work, which essentially aims to evaluate our biomechanical model, we have adopted simple tongue muscle activation functions, linear in pieces, each piece being defined by its duration and by activation levels $\alpha$ at the start (initial) and at the target (final) of the movement (Figure 4.4). In the simulations presented in

Figure 4.4, each muscle is activated in 100 ms with an activation maintenance of up to 350 ms (in red in Figure 4.4).

### 4.5.3. *Tongue shifts*

Figure 4.5 shows the movements (in meter) of the tongue as a result of the independent activation of the different lingual muscles. The order of magnitude of the movements and the final forms of the tongue chime with the information available in the literature (Buchaillard *et al.* 2009; Fang *et al.* 2009; Hermant *et al.* 2017).

### 4.5.4. *Tongue trajectories*

Because of the very large deformations it can undergo (Napadow *et al.* 1999), the tongue is capable of producing complex trajectories.

To illustrate our point, Figures 4.6 and 4.7 show respectively the trajectory and temporal displacement of two points (located respectively in the palatal area and in the pharyngeal zone of the tongue) resulting from the successive activation of the posterior genioglossus (GGP) and hyoglossus (HG) muscles. These two muscles were chosen because they play a major role in the production of phonemes /i/ and /a/. Each muscle is activated for 100 ms, with a maintenance period of 200 ms and a descent period in 100 ms, then 200 ms for return to a stable position. We chose relatively long durations compared to speech or swallowing movements (a transition between two phonemes lasts about a few tens of ms), in order to ensure that the effects of muscle activations are fully visible and to avoid any overlapping of the effects of each activation that could be generated by a rapid sequence of movements. We find the characteristic effects of each muscle:

1) a strong advance of the tongue in the pharyngeal part due to the activation of the genioglossus, accompanied by an elevation of the tongue in the palatal part;

2) a lowering of the tongue related to the action of the hyoglossus muscle.

The trajectories are curved, as we observe in the data measured by articulography on human subjects. We also note that in addition to the average path of the trajectory, small oscillations are superimposed which are probably due to the relative length of the movements, on the scale of the movements usually observed in speech.

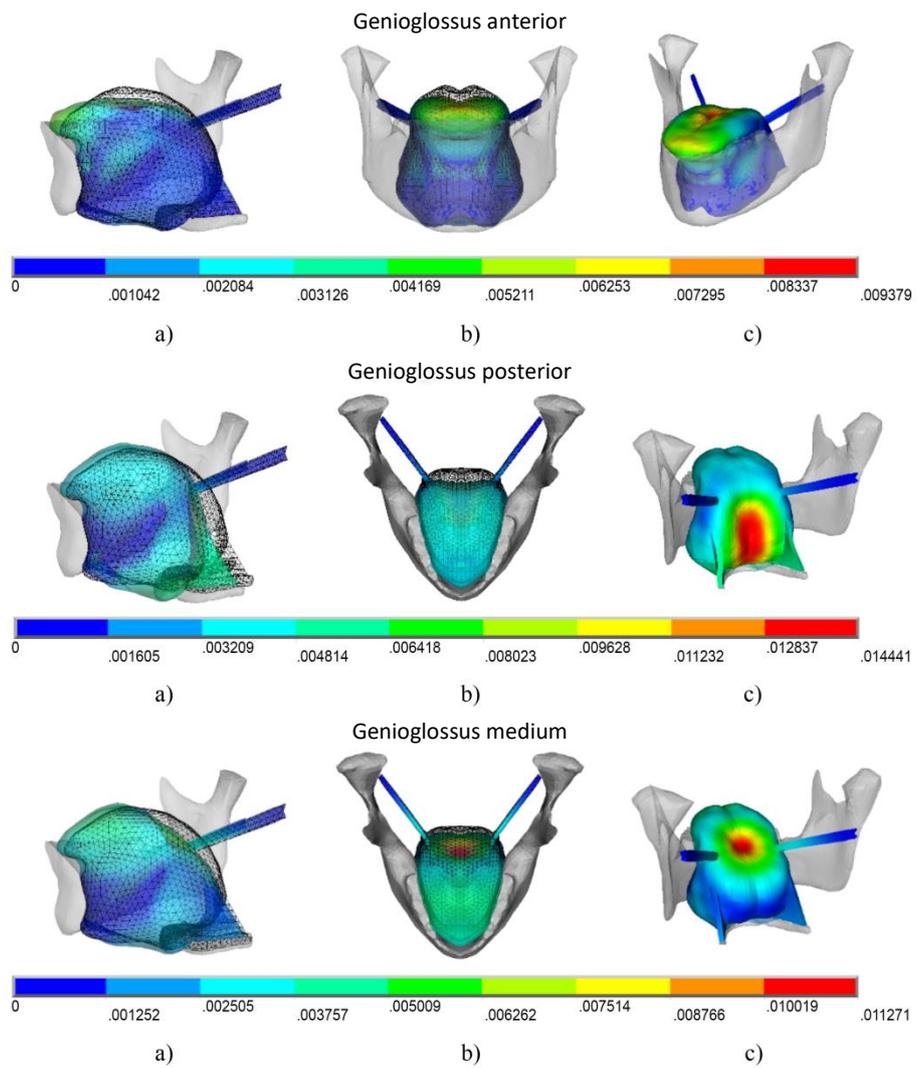

**Figure 4.5A.** *Displacements of the tongue due to individual activation independent of the tongue muscles. The color map represents the total displacement level (in meter). The transparent mesh represents the initial configuration of the mesh. a) Top view; (b) Front view; (c) Cavalier perspective (continued).*

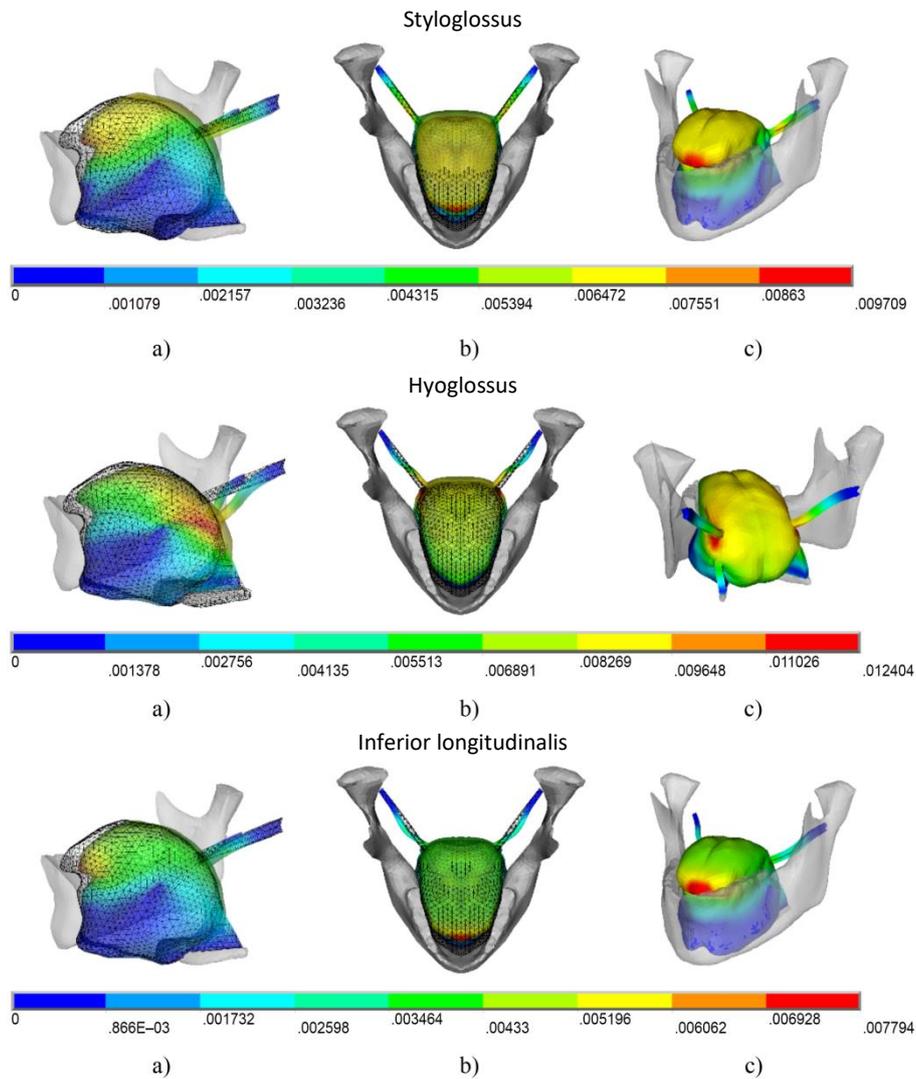

**Figure 4.5B.** *Displacements of the tongue due to individual activation independent of the tongue muscles. The color map represents the total displacement level (in meter). The transparent mesh represents the initial configuration of the mesh. a) Top view; (b) Front view; (c) Cavalier perspective (continued).*

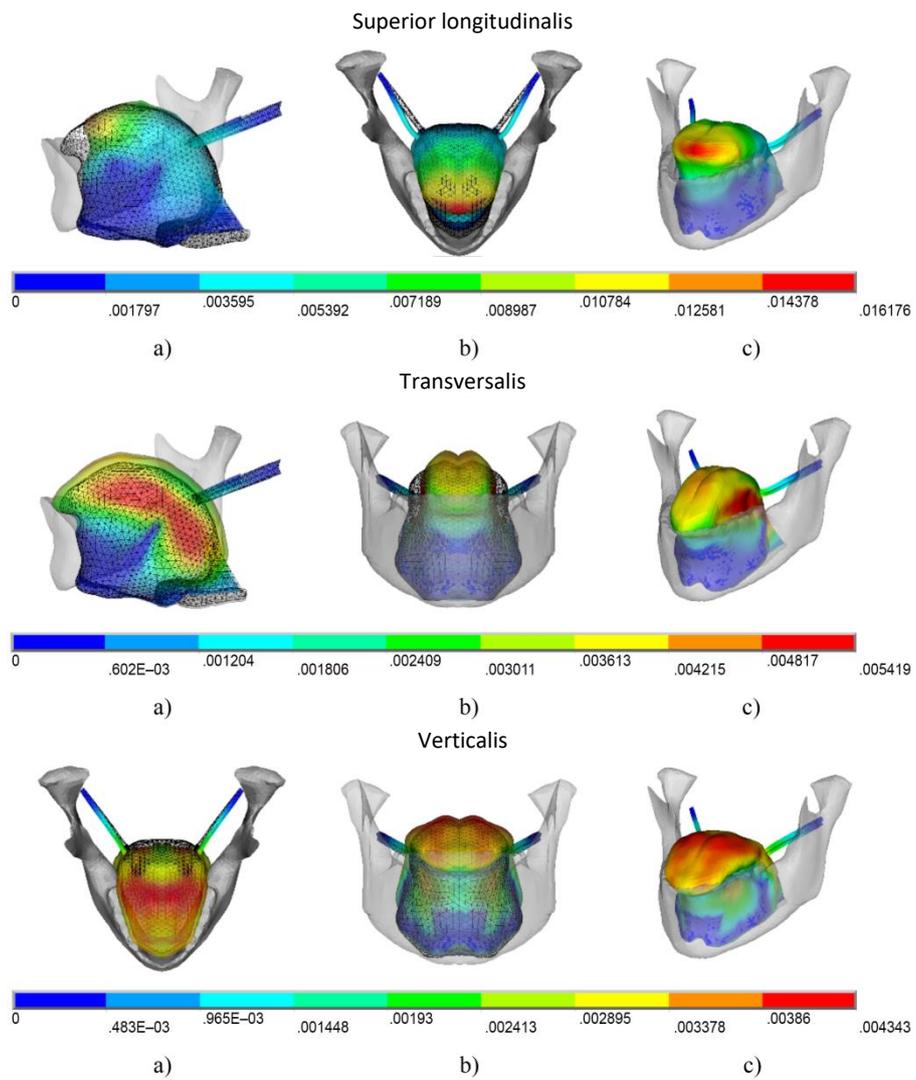

**Figure 4.5C.** *Displacements of the tongue due to individual activation independent of the tongue muscles. The color map represents the total displacement level (in meter). The transparent mesh represents the initial configuration of the mesh. a) Top view; (b) Front view; (c) Cavalier perspective.*

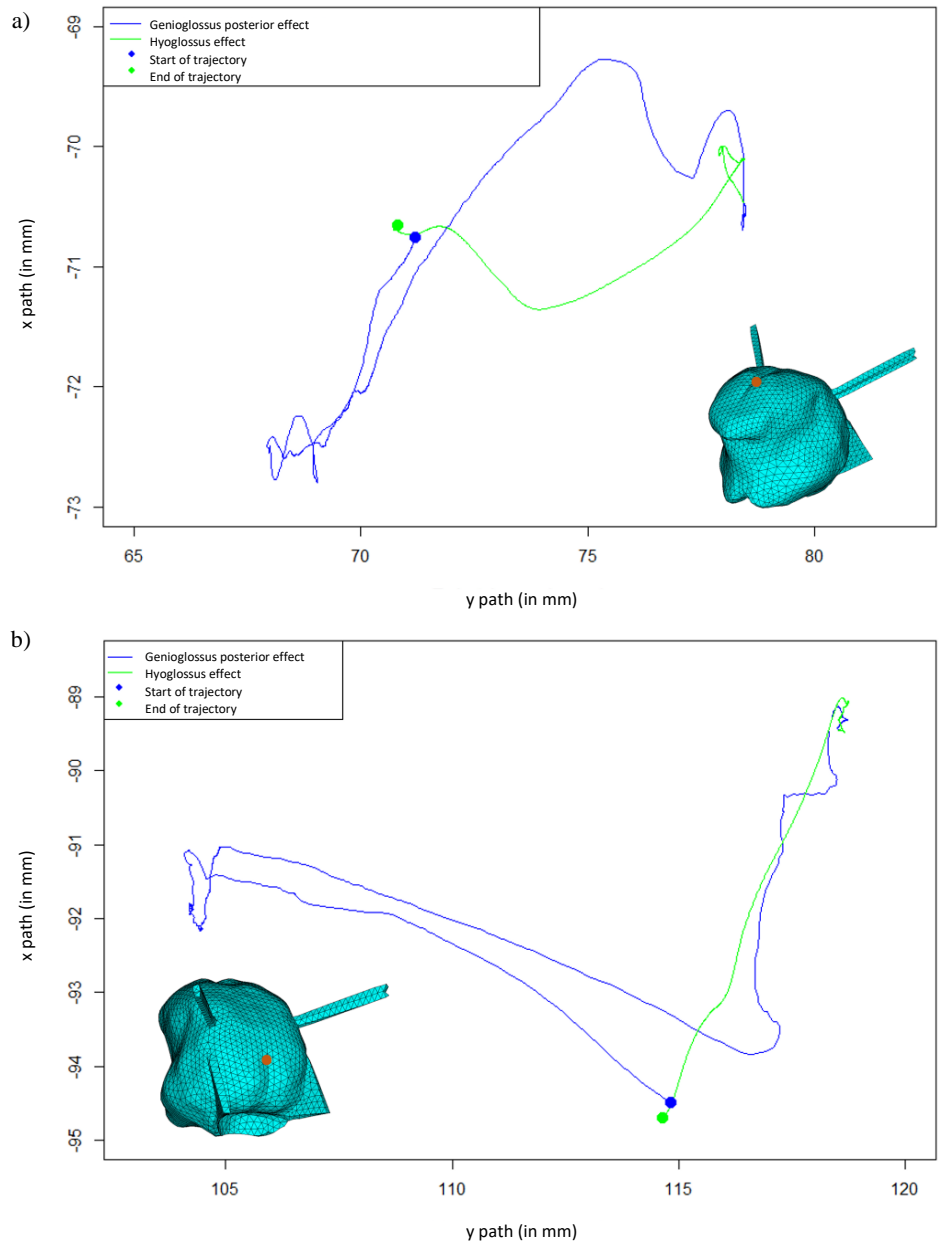

**Figure 4.6.** *Trajectory of two points in the median sagittal plane, located respectively in the dorsal zone (a) and in the pharyngeal zone (b) of the tongue, resulting from the successive activations of the GGP, then of the HG, two major muscles in the production of /i/ and /a/.*

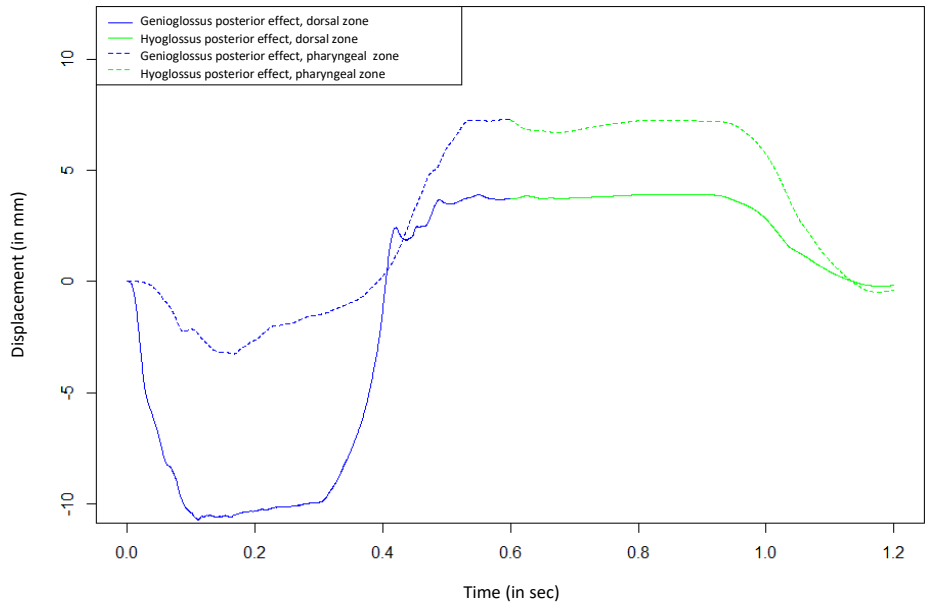

**Figure 4.7.** *Horizontal temporal displacement of two points, located respectively in the dorsal and pharyngeal zone of the tongue resulting from a simulation acting successively the GGP, then the HG, two major muscles in the production of /i/ and /a/.*

## 4.6. Discussion

The application framework presented is not common in the field of biomechanics. Indeed, the activation times of the muscles are of the order of a few tens of milliseconds (between 50 ms and 100 ms), which induces strong accelerations of the system and strong inertia effects. The challenge of our modeling is not only to describe the tongue deformations generated by muscle activations, but to correctly describe the trajectories passing from the initial form to the final form. It is these trajectories that shape the sounds of speech and influence their perception. It is the temporal characteristics of lingual movements that guarantee the efficiency of swallowing.

As illustrated by the results presented in section 4.5.4, setting in motion a biomechanical system such as the tongue, following muscle activations in ramps (Figure 4.4), generates complex trajectories with varyingly rapid transitions between quasi-static positions and small local oscillations. This complexity can be amplified in the case of activations of several muscles successively (Figure 4.7).

The other specificity of the modeling of the human tongue is the very large deformations that this organ can undergo during functions such as swallowing and phonation. It is interesting, from this point of view, to note that levels of Von Mises deformations of the order of 50% (Hencky strain calculated by Ansys Mechanical APDL®) are observed for all muscle activations presented in Figure 4.5. This suggests even higher levels of deformity for extreme gestures, such as when the tongue is pulled out of the mouth (Napadow *et al.* 1999).

## 4.7. Perspective: model order reduction for real-time simulation

The question of computational times for a biomechanical model like ours is important in numerical simulation, which can make the incorporation of such models into clinical application frameworks prohibitive. For example, the 350 ms sequences of lingual movements, the simulation results of which are shown in Figure 4.5, required just over 5 hours of machine computing each (on a machine equipped with ten cores with a 3.31 GHz processor). One of the answers given in the literature is the notion of *model order reduction*. It is a numerical simulation method that integrates model physics, while reducing computational complexity. Most of the model reduction methods presented in the literature are applicable in the context of static or quasi-static models. In the case of dynamic models, Ansys® has developed a method that takes this property into account in a non-linear framework.

This method is based on the use of machine learning methods adapted to time series, such as recurrent neural networks (Rumelhart *et al.* 1986). Eventually, a software tool integrating a scale model of the biomechanical model of the tongue should emerge, in order to synthesize sequences of movement of the tongue in cases of natural and pathological speech in real time (Calka *et al.* 2021). This will also allow, in a second step, for the creation of a tool to simulate the ability to move the tongue after surgery.

## 4.8. Conclusion

The last thirty years of study and biomechanical modeling of the tongue have made it possible to note the difficulty of this task, both when it comes to modeling (non-linearities of tissues, anisotropies, visco-elasticities) and numerical simulation (transient analysis, large deformations, calculation time).

In this chapter, the most exhaustive description possible of our biomechanical model in comparison with the literature shows the progress made to arrive at anatomical fidelity of the subject and the muscular structures of the tongue, as well as the constitutive laws.

In addition, our real-time simulation work should, in the short and medium term, be able to be used in applications for speech production and tongue pathologies.

Finally, recent MRI data from our reference subject during phonation tasks will make it possible to precisely evaluate our model, in order to validate it in the application contexts mentioned above.

## 4.9. Bibliography

Acland, R.D. (2003). Acland's video atlas of human anatomy. Video, Lippincott Williams & Wilkins.

Anderson, P., Fels, S., Stavness, I., Pearson Jr., W.G., Gick, B. (2019). Intravelar and ex-ravelar portions of soft palate muscles in velic constrictions: A three-dimensional modeling study. *Journal of Speech, Language, and Hearing Research*, 62(4), 802–814.

Buchaillard, S., Brix, M., Perrier, P., Payan, Y. (2007). Simulations of the consequences of tongue surgery on tongue mobility: implications for speech production in post-surgery conditions. *The International Journal of Medical Robotics and Computer Assisted Surgery*, 3(3), 252–261.

Buchaillard, S., Perrier, P., Payan, Y. (2009). A biomechanical model of cardinal vowel production: Muscle activations and the impact of gravity on tongue positioning. *The Journal of the Acoustical Society of America*, 126(4), 2033–2051.

Calka, M., Perrier, P., Ohayon, J., Grivot-Boichon, C., Rochette, M., Payan, Y. (2021). Machine-Learning based model order reduction of a biomechanical model of the human tongue. *Computer Methods and Programs in Biomedicine*, 198, 105786.

Fang, Q., Fujita, S., Lu, X., Dang, J. (2009). A model-based investigation of activations of the tongue muscles in vowel production. *Acoustical Science and Technology*, 30(4), 277–287.

Feldman, A.G. (1986). Once more on the equilibrium-point hypothesis (λ model) for motor control. *Journal of motor behavior*, 18(1), 17–54.

Fujita, S., Dang, J., Suzuki, N., Honda, K. (2007). A computational tongue model and its clinical application. *Oral Science International*, 4(2), 97–109.

Gerard, J.-M., Ohayon, J., Luboz, V., Perrier, P., Payan, Y. (2005). Non-linear elastic properties of the lingual and facial tissues assessed by indentation technique: Application to the biomechanics of speech production. *Medical engineering & physics*, 27(10), 884–892.

Gerard, J.-M., Wilhelms-Tricarico, R., Perrier, P., Payan, Y. (2006). A 3D dynamical biomechanical tongue model to study speech motor control [Online]. Available from: https://arxiv.org/abs/physics/0606148v1.

Gick, B., Allen, B., Roewer-Després, F., Stavness, I. (2017). Speaking tongues are actively braced. *Journal of Speech, Language, and Hearing Research*, 60(3), 494–506.

Gomi, H., Kawato, M. (1996). Equilibrium-point control hypothesis examined by measured arm stiffness during multijoint movement. *Science*, 272(5258), 117–120.

Gray, H. (1918). Anatomy of the human body. *Annals of surgery*, 68(5), 564–566.

Gribble, P.L., Ostry, D.J., Sanguineti, V., Laboissière, R. (1998). Are complex control signals required for human arm movement?. *Journal of neurophysiology*, 79(3), 1409–1424.

Huxley, A.F. (1957). Muscle structure and theories of contraction. *Prog. Biophys. Biophys. Chem*, 7, 255–318.

Kakita, Y., Fujimura, O. (1977). Computational model of the tongue: A revised version. *The Journal of the Acoustical Society of America*, 62(S1), S15–S16.

Kappert, K.D.R., van Alphen, M.J.A., van Dijk, S., Smeele, L.E., Balm, A.J.M., van der Heijden, F. (2019). An interactive surgical simulation tool to assess the consequences of a partial glossectomy on a biomechanical model of the tongue. *Computer methods in biomechanics and biomedical engineering*, 22(8), 827–839.

Kappert, K.D.R., Connesson, N., Elahi, S.A., Boonstra, S., Balm, A.J.M., van der Heijden, F., Payan, Y. (2021). In-vivo tongue stiffness measured by aspiration: Resting vs general anesthesia. *Journal of biomechanics*, 114, 110147.

Napadow, V.J., Chen, Q., Wedeen, V.J., Gilbert, R.J. (1999). Intramural mechanics of the human tongue in association with physiological deformations. *Journal of biomechanics*, 32(1), 1–12.

Patri, J.-F., Diard, J., Perrier, P. (2015). Optimal speech motor control and token-to-token variability: a Bayesian modeling approach. *Biological cybernetics*, 109(6), 611–626.

Payan, Y., Perrier, P. (1997). Synthesis of VV sequences with a 2D biomechanical tongue model controlled by the Equilibrium Point Hypothesis. *Speech communication*, 22(2-3), 185–205.

Pelteret, J.-P., Reddy, B.D. (2012). Computational model of soft tissues in the human upper airway. *International journal for numerical methods in biomedical engineering*, 28(1), 111–132.

Perrier, P., Fuchs, S. (2008). Speed–curvature relations in speech production challenge the 1/3 power law. *Journal of neurophysiology*, 100(3), 1171–1183.

Perrier, P., Payan, Y., Zandipour, M., Perkell, J. (2003). Influences of tongue biomechanics on speech movements during the production of velar stop consonants: A mode-ling study. *The Journal of the Acoustical Society of America*, 114(3), 1582–1599.

Rodrigues, M.A.F., Gillies, D., Charters, P. (2001). A biomechanical model of the upper airways for simulating laryngoscopy. *Computer Methods in Biomechanics and Biomedical Engineering*, 4(2), 127–148.

Rohan, P.-Y., Lobos, C., Nazari, M.A., Perrier, P., Payan, Y. (2017). Finite element models of the human tongue: a mixed-element mesh approach. *Computer Methods in Biomechanics and Biomedical Engineering: Imaging & Visualization*, 5(6), 390–400.

Rumelhart, D.E., Hinton, G.E., Williams, R.J. (1986). Learning representations by back-propagating errors. *Nature*, 323(6088), 533–536.

Schiavone, P., Boudou, T., Promayon, E., Perrier, P., Payan, Y. (2008). A light sterilizable pipette device for the in vivo estimation of human soft tissues constitutive laws. Dans *30th Annual International Conference of the IEEE Engineering in Medicine and Biology Society*. IEEE, 4298–4301.

Stavness, I., Lloyd, J.E., Payan, Y., Fels, S. (2011). Coupled hard–soft tissue simulation with contact and constraints applied to jaw–tongue–hyoid dynamics. *International Journal for Numerical Methods in Biomedical Engineering*, 27(3), 367–390.

Stavness, I., Gick, B., Derrick, D., Fels, S. (2012). Biomechanical modeling of English/r/ variants. *The Journal of the Acoustical Society of America*, 131(5), EL355–EL360.

Wang, Y.K., Nash, M.P., Pullan, A.J., Kieser, J.A., Röhrle, O. (2013). Model-based identification of motion sensor placement for tracking retraction and elongation of the tongue. *Biomechanics and modeling in mechanobiology*, 12(2), 383–399.

Wilhelms-Tricarico, R. (1995). Physiological modeling of speech production: Methods for modeling soft-tissue articulators. *The Journal of the Acoustical Society of America*, 97(5), 3085–3098.

Winters, J.M. (1990). Hill-based muscle models: a systems engineering perspective. Dans *Multiple muscle systems*, Winters, J.M, Woo, S.L.Y. (dir.). Springer, Berlin, 69–93.

Yousefi, A.-A.K., Nazari, M.A., Perrier, P., Panahi, M.S., Payan, Y. (2018). A visco-hyperelastic constitutive model and its application in bovine tongue tissue. *Journal of biomechanics*, 71, 190–198.